\documentclass[
 reprint,
superscriptaddress,
 amsmath,amssymb,
 aps,
superscriptaddress,
 amsmath,amssymb,
 aps,longbibliography
]{revtex4-1}

\usepackage{graphicx}
\usepackage{dcolumn}
\usepackage{bm}
\usepackage{amsmath,amssymb,amstext,amsthm}
\usepackage{braket}
\usepackage{bbold}
\usepackage{bbm}
\usepackage{xcolor}
\usepackage{ gensymb }
\usepackage{soul}
\setstcolor{red}
\usepackage{ mathrsfs }
\usepackage{cancel}

\usepackage{hyperref}

\hypersetup{colorlinks,linkcolor=blue,citecolor=blue,urlcolor=blue}

\newcommand{\HE}[0]{$^3\text{He}$}

\begin{document}

\preprint{APS/123-QED}
\setlength{\abovedisplayskip}{1pt}
\title{Multimode structured neutron beams}

\author{Owen Lailey} 
\email{oalailey@uwaterloo.ca}
\affiliation{Institute for Quantum Computing, University of Waterloo,  Waterloo, ON, Canada, N2L3G1}
\affiliation{Department of Physics and Astronomy, University of Waterloo, Waterloo, ON, Canada, N2L3G1}

\author{Dusan Sarenac}
\affiliation{Department of Physics, University at Buffalo, State University of New York, Buffalo, New York 14260, USA}

\author{Charles W. Clark}
\affiliation{Joint Quantum Institute, National Institute of Standards and Technology and University of Maryland, College Park, Maryland 20742, USA}

\author{David G. Cory}
\affiliation{Institute for Quantum Computing, University of Waterloo,  Waterloo, ON, Canada, N2L3G1}
\affiliation{Department of Chemistry, University of Waterloo, Waterloo, ON, Canada, N2L3G1}

\author{Lisa DeBeer-Schmitt} 
\affiliation{Neutron Scattering Division, Oak Ridge National Laboratory, Oak Ridge, TN 37831, USA}

\author{Huseyin Ekinci} 
\affiliation{Institute for Quantum Computing, University of Waterloo,  Waterloo, ON, Canada, N2L3G1}
\affiliation{Department of Physics and Astronomy, University of Waterloo, Waterloo, ON, Canada, N2L3G1}

\author{Davis V. Garrad} 
\affiliation{Department of Physics and Astronomy, University of Waterloo, Waterloo, ON, Canada, N2L3G1}

\author{Melissa E. Henderson} 
\affiliation{Neutron Scattering Division, Oak Ridge National Laboratory, Oak Ridge, TN 37831, USA}

\author{Michael G. Huber}
\affiliation{National Institute of Standards and Technology, Gaithersburg, Maryland 20899, USA}

\author{Priyanka Vadnere}
\affiliation{Department of Physics, University at Buffalo, State University of New York, Buffalo, New York 14260, USA}

\author{Kirill Zhernenkov}
\affiliation{J\"ulich Centre for Neutron Science at Heinz Maier-Leibnitz Zentrum, Forschungszentrum J\"ulich GmbH, 85748 Garching, Germany}

\author{Dmitry A. Pushin}
\email{dmitry.pushin@uwaterloo.ca}
\affiliation{Institute for Quantum Computing, University of Waterloo,  Waterloo, ON, Canada, N2L3G1}
\affiliation{Department of Physics and Astronomy, University of Waterloo, Waterloo, ON, Canada, N2L3G1}

\date{\today}

\pacs{Valid PACS appear here}

\begin{abstract}
The experimental realization of neutron orbital angular momentum (OAM) states and neutron Airy beams has opened new avenues for structured neutron science in both materials characterization and fundamental physics. These additional degrees of freedom in scattering experiments enable the exploration of selection rules for neutrons, the analysis of scattering properties in topological materials, and the generation of auto-focusing neutron beams. In the effort to enhance the amount of spatial and angular-momentum information retrievable from a single measurement, and to overcome current phase-grating efficiency limits, here we demonstrate multimode structured neutron beams that enable simultaneous access to multiple, well-defined OAM modes, and to hybrid combinations of OAM and Airy states. This multimode approach, analogous to wavelength- or OAM-multiplexing in optics, facilitates the efficient investigation of material scattering properties and nuclear interactions with a neutron source composed of a discretized OAM spectrum.

\end{abstract}
\maketitle

\section{Introduction}
Structured light and matter waves have enabled diverse applications in imaging, communication, quantum information processing, and precision measurement~\cite{rubinsztein2016roadmap, bliokh2023roadmap}. Beyond generating a single structured state, such as an orbital angular momentum (OAM) mode, optical multiplexing techniques have developed to exploit multiple co-propagating OAM modes to enhance information capacity and parallelism in data transmission and imaging~\cite{wang2012terabit, bozinovic2013terabit}. In matter wave experiments, such as with neutrons, a similar form of multiplexing exists in the wavelength domain: the broad wavelength distribution can be effectively demultiplexed through time-of-flight detection, allowing the experimental signal to be analyzed as separate spectral components~\cite{copley1993neutron, unruh2007high}. The analogous implementation of multiplexed structured matter waves, in which multiple well-defined OAM or other structured states are accessed simultaneously, remains largely unexplored.

Recently, neutron OAM states have been experimentally realized by transmitting neutrons through nanofabricated silicon phase-grating arrays composed of millions of micron-scale $q$-fold fork-dislocations, generating neutrons with OAM number $\ell = mq$ in the $m$th diffraction order ~\cite{sarenac2022experimental}. These demonstrated structured neutron waves enable a wide range of proposed experiments, including the study of selection rules for neutrons with polarized \HE ~\cite{jach2022method}, the analysis of scattering in topological materials using helical waves~\cite{henderson2021characterization, henderson2022skyrmion, henderson2023three, henderson2024quantum} and the investigation of angular momentum-dependent nuclear scattering processes~\cite{karlovets2015, larocque2018twisting, afanasev2019schwinger, afanasev2021elastic, sherwin2022scattering, pavlov2025angular}. However, several experimental challenges persist, including the sub-micron alignment precision required for producing coherent superpositions of neutron OAM modes $\ell = \pm 3$ demonstrated in Ref.~\cite{sarenac2024small}, and the currently low diffraction efficiency of individual phase-gratings. To improve experimental flexibility and to enhance the information capacity of a single measurement, multiple phase-gratings can be stacked in sequence, enabling simultaneous access to several distinct OAM modes, or even combined Airy~\cite{sarenac2025generation} and helical states.

In this work, we experimentally generate and characterize multimode structured neutron beams, composed of combinations of $\ell = 3$ and $\ell=7$ OAM modes and hybrid configurations of Airy and helical states. Using small angle neutron scattering (SANS) measurements, we map the resulting far-field intensity distributions of the multiplexed structured neutron beam at a position-sensitive neutron detector and observe excellent agreement with simulations. These results pave the way for neutron OAM spectroscopy measurements, analogous to developments with light and electrons~\cite{xie2017using, berger2018spectroscopy, grillo2017measuring}, and for two-dimensional (2D) neutron multiplexing techniques that provide both angular-momentum and energy resolution in a single measurement.

\begin{figure}
    \centering\includegraphics[width=1\linewidth]{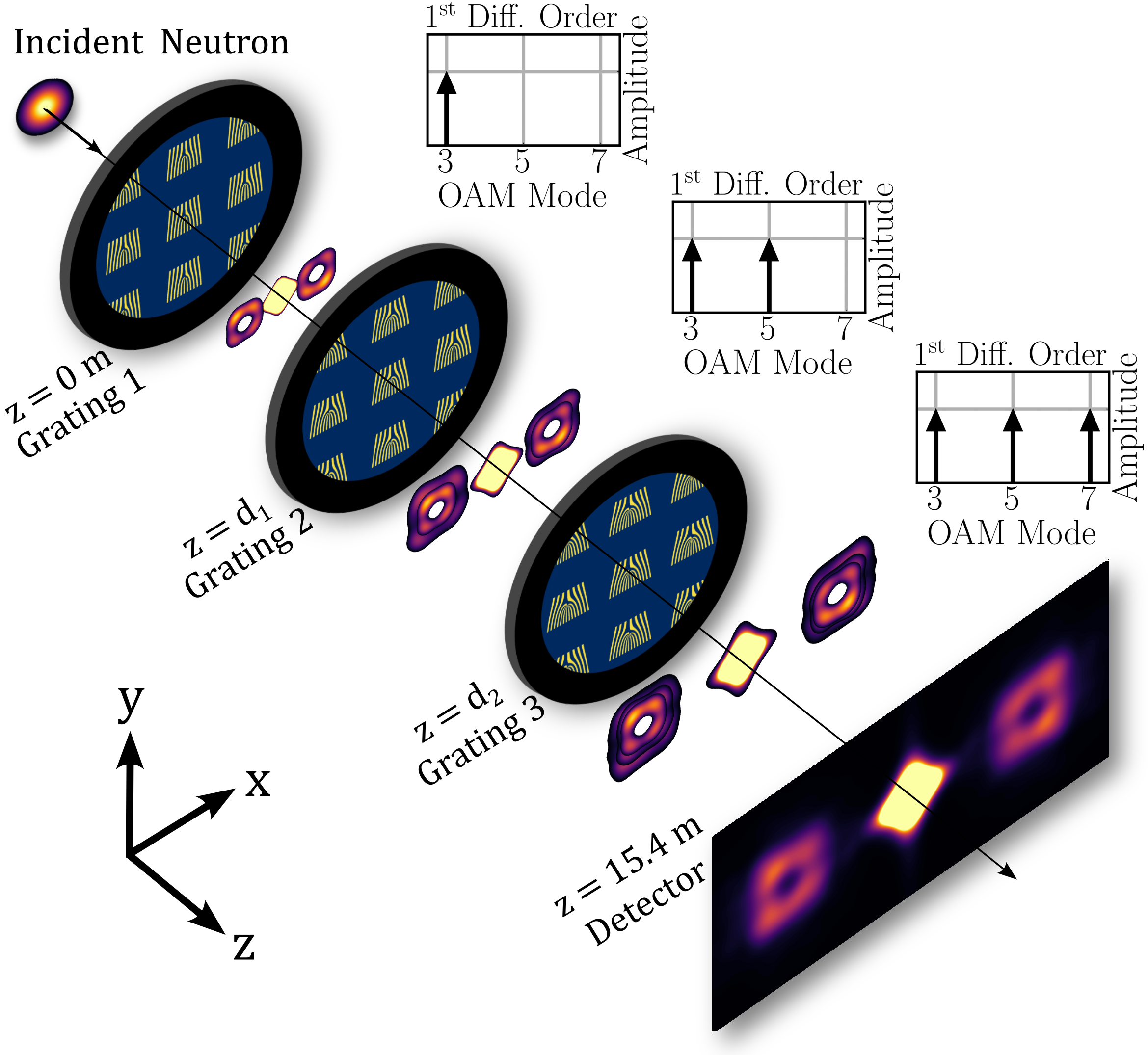}
    \caption{Schematic illustration of multimode structured neutron beam generation. An incident neutron beam passes through a series of phase-grating arrays, spatially separated along the beam (z) axis at distances $d_1$, $d_2$, etc. In this illustration, each grating array contains fork dislocations with topological charges $q=3,~5,$ or $7$, diffracting neutrons into well-defined OAM states in the first diffraction orders. After the final grating (Grating 3), the beam consists of a mixture of three distinct OAM states, which co-propagate to the detector plane. The resulting intensity profile exhibits a broadened doughnut-shaped structure, as shown in the simulated diffraction pattern. All gratings share the same periodicity and the separation between the gratings is on the order of a few mm. The size of the grating arrays within the mounts is exaggerated for visual clarity.}
    \label{fig:schematic}
\end{figure}

\section{Materials and Methods}
We implemented three different silicon wafers, each patterned with an array of phase-gratings etched on one face.  Two of the gratings were fork-dislocation phase-gratings, one with topological charge $q = 3$, another with $q = 7$, and the third was a cubic phase-grating. The profile of an individual fork-dislocation phase-grating in the array is given by:

\begin{align}
	\frac{1}{2}\text{sgn}\left(\text{sin}\left[\frac{2\pi  }{p}x+q\tan^{-1}(y/x)\right]\right)+\frac{1}{2},
	\label{eqn:gratingProfile}
\end{align}
\noindent where $p$ is the grating period and $x$,~$y$ are the transverse coordinates. Likewise, the profile of an individual cubic phase-grating in the array is given by:

\begin{equation}
    \frac{1}{2}  \text{sgn} \left(\sin \left[ \frac{2\pi}{p}x + c_xx^3 - c_yy^3  \right] \right) + \frac{1}{2} ,
    \label{eq:phase_prof}
\end{equation}

\noindent where $c_x,~c_y$ are the cubic coefficients of the grating. Each array covered a 0.5 cm by 0.5 cm area and consisted of $6,250,000$ individual 1 $\mu$m by 1 $\mu$m phase-gratings.  Each grating had a period of $120$ nm and was separated by 1 $\mu$m from adjacent gratings on all sides. The grating heights for the $q=3,~7$, and cubic gratings were $500$~nm, $400$~nm, and $300$~nm respectively. The fabrication procedure and Scanning Electron Microscope (SEM) images for the fork-dislocation phase-gratings are available in Ref.~\cite{sarenac2022experimental} and likewise for the cubic phase-gratings in Ref.~\cite{sarenac2025generation}. 

The experiments were performed at the GP-SANS beamline at the High Flux Isotope Reactor at Oak Ridge National Laboratory~\cite{wignall201240}. Experimental parameters were similar to those in Ref.~\cite{sarenac2024small}, which described the interference of neutron helical waves. The wafers were mounted on rotation stages positioned 17.8 m from a 20 mm diameter source aperture. A 4 mm diameter sample aperture was placed directly in front of the gratings. The detector was located 15.4 m downstream of the sample, with the detector spanning an area of approximately $1$ m$^2$ and individual pixel dimensions of approximately $5.5$ mm by $4.3$ mm.  The neutron wavelength distribution was triangular with $\Delta\lambda/\lambda\approx0.13$, where $\Delta\lambda$ is the FWHM and the central wavelength is $12$~\AA. Wavelength selection was achieved using a turbine-like velocity selector, which eliminates fractional $\lambda$ contributions typically present with monochromator crystals.

To improve environmental isolation from the experimental hall, a $\approx 33 \times 30 \times 38$~cm $(l\times w \times h)$ aluminum enclosure was placed over the gratings. A 5-cm hole was cut in the front face adjacent to the sample aperture, and a 15-cm hole in the back face allowed diffracted neutrons to pass through to the detector 15.4 m downstream.

We collected an empty beam scan (without a sample) and a background scan using plain silicon wafers of equivalent size and thickness. These measurements were used to take into account the factors that contribute to losses in intensity and increased background scattering noise. Final SANS data were processed using a low-pass filter to reduce pixel-to-pixel Poissonian noise.

\begin{figure*}
    \centering\includegraphics[width=1\linewidth]{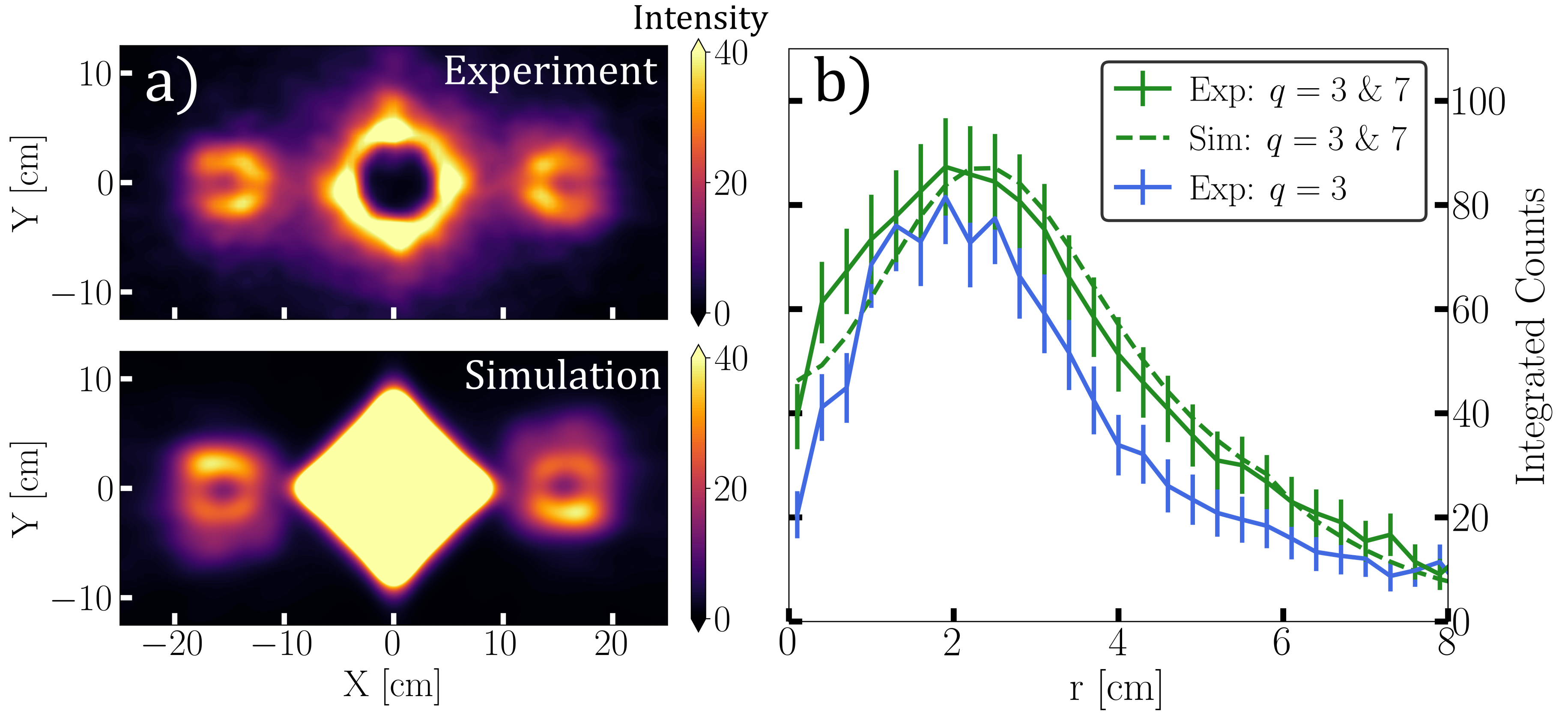}
    \caption{a) The measured and simulated diffraction spectra at $z = 15.4$~m after a combination of a $q=3$ and a $q=7$ fork dislocation phase-grating. The smaller $q = 3$ doughnut-shaped profile is clearly observed and partially surrounded by the larger, less intense $q = 7$ profile. The $q=7$ grating is rotated 4 degrees with respect to the $q = 3$ phase-grating as seen in the experiment and simulation. b) The azimuthally integrated profile of the $m=+1$ diffraction order from the $q = 3$ and $q=7$ diffraction gratings in a). To compare, we measured a reference $q = 3$, $m = +1$ profile (analogous to Ref.~\cite{sarenac2022experimental}) to distinguish the contribution from the $q = 7$ grating. Notably, there is increased intensity at larger radial distances $r$ due to the larger $q = 7$ doughnut profile. The two measurements were normalized to the same measurement time. In the experiments a beam trap was placed at the center of the detector in the direct beam path to better emphasize neutron counts of the nonzero diffraction orders at the detector.}
    \label{fig:oam}
\end{figure*}

\begin{figure*}
    \centering\includegraphics[width=1\linewidth]{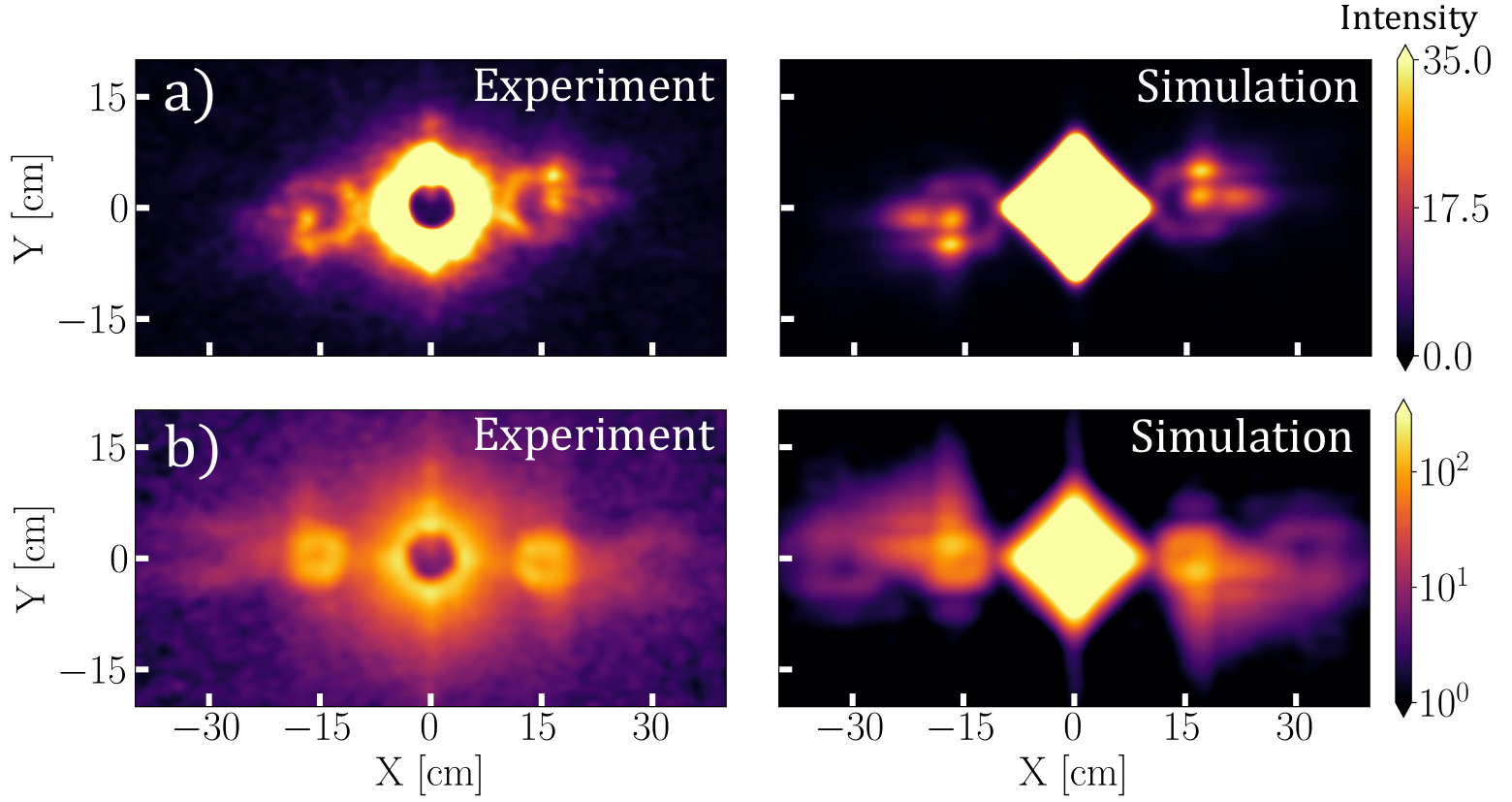}
    \caption{a) The measured and simulated diffraction spectra at $z = 15.4$~m after a cubic phase-grating and a $q=7$ fork dislocation phase-grating. The $m = \pm 1$ diffraction orders show a superposition of the Airy beam profile and the $q=7$ doughnut-shaped profile. b) The measured and simulated diffraction spectra at $z = 15.4$~m after a cubic phase-grating and a $q = 3$ fork dislocation phase-grating. Here we plot the measured intensity in log scale to better emphasize the Airy profile as well as the second diffraction order. In the experiments a beam trap was placed at the center of the detector in the direct beam path to better emphasize neutron counts of the nonzero diffraction orders at the detector.}
    \label{fig:air_oam}
\end{figure*}

\section{Results}
A general schematic for multimode structured neutron beam generation is illustrated in Fig.~\ref{fig:schematic}. We experimentally implemented several combinations of spatially separated phase-grating arrays and measured the far-field intensity profiles. First, we generated an input state composed of two distinct neutron OAM beams by placing the $q = 3$ grating at $z = 0$ and aligning the $q=7$ downstream at $d_1 = 3.5$~mm $\pm~0.1$~mm. Shown in Fig.~\ref{fig:oam}a are the measured and simulated intensity profiles at $z = 15.4$~m displaying the combination of the more compact $q = 3$ profile and the broader $q = 7$ intensity profile. To simulate the experiment, we employed the structured neutron SANS model developed in~\cite{sarenac2024small, sarenac2025generation}, based on the Fresnel-Kirchoff diffraction integral. We propagate a Gaussian neutron wave-packet through the diffraction grating arrays and $15.4$~m downstream to the detector. 

In Fig.~\ref{fig:oam}b we plot the azimuthally integrated intensity profile of the $m = +1$ diffraction order for the $q = 3~\&~7$ input state and compare it with a reference measurement from a single $q = 3$ grating. The two different measurements are normalized to the same measurement time. The contribution of the larger $q=7$ mode is evident as an increased intensity across the radial distances $r$, relative to the $q=3$ case, indicating that these two input beams are distinct.

These techniques are not limited to OAM modes alone; they can also accommodate other structured neutron states, such as Airy beams. To experimentally demonstrate this, we placed a cubic phase-grating at $z = 0$ and positioned a $q = 7$ fork dislocation grating downstream at $d_1 = 3.3$~mm $\pm~0.1$~mm. Shown in Fig.~\ref{fig:air_oam}a are the measured and simulated intensity patterns of the superimposed Airy profile and $q=7$ doughnut-shaped profile in the first diffracted orders at the position sensitive neutron detector. 
In the experiment, the $q=7$ grating was deliberately rotated by several degrees to align the doughnut-shaped profile with the second lobe of the Airy beam. This rotation was also included in the simulation, where the $q=7$ grating was rotated 4 degrees counter clockwise relative to the cubic phase-grating. 

Shown in  Fig.~\ref{fig:air_oam}b are the measured diffraction spectra displaying the linear superposition of the neutron Airy beam and the $q=3$ intensity profile in the nonzero diffraction orders. We placed the $q=3$ grating at $z = 0$ and positioned the cubic grating behind it at $d_1 = 3.0$~mm $\pm~0.1$~mm. In this case, we plot the logarithm of the measured intensity to highlight the Airy profile relative to the more intense $q=3$ profile, and to visualize the less intense second diffraction order. The $q=3$ intensity profiles dominate in both the first and second diffracted orders, in good agreement with simulation, since they span a smaller spatial area and are produced from a significantly taller grating than the Airy profiles.

\section{Discussion and Conclusion}
We have experimentally realized multimode structured neutron beams, composed of orthogonal helical modes and hybrid Airy–OAM beams, enabling simultaneous measurements with well-defined structured neutrons. The incorporation of multiple spatially separated phase-gratings transforms the limitations of low-efficiency gratings into a useful beam property, allowing the input beam to consist of a tailored spectrum of structured states. This approach offers a flexible and scalable path toward implementing complex neutron wavefronts in scattering experiments where single-mode generation or precise grating alignment would otherwise be prohibitive. Moreover, this work adds a new modality to the structured neutron toolbox~\cite{clark2015controlling, nsofini2016spin, sarenac2018methods, sarenac2019generation, geerits2021twisting, sarenac2022experimental, sarenac2023novel, geerits2023phase, le2023spin, sarenac2024small, sarenac2025generation, pushin2025advancements, geerits2025measuring}, expanding the range of accessible wavefronts for probing materials and fundamental interactions~\cite{karlovets2015, larocque2018twisting, afanasev2019schwinger, afanasev2021elastic, sherwin2022scattering, jach2022method, pavlov2025angular}.

This proof-of-principle experiment enables several promising applications. First, by utilizing multimode structured neutron beams, we can realize neutron OAM spectroscopy measurements, analogous to similar techniques applied to light and electrons~\cite{xie2017using, berger2018spectroscopy, grillo2017measuring}. This would allow for the complete characterization of the OAM spectrum both before and after scattering, providing insights into whether specific OAM modes are preferentially scattered or absorbed by the sample. Additionally, the use of Airy beams, known for their properties of nondiffraction and self-healing, could reveal unique scattering behaviors when compared to conventional $\ell=0$ or $\ell\neq0$ reference beams. Furthermore, this work paves the way for 2D neutron wave multiplexing: OAM and energy. By implementing multimode structured neutron beams at a pulsed neutron source, a single measurement obtains both energy resolved and angular momentum resolved scattering data. Finally, even when considering a single OAM mode, stacking identical gratings enhances the intensity of the first diffracted orders, providing a feasible alternative to fabricating higher-aspect-ratio phase-gratings.

As an extension of this technique, parallel stacking of phase-gratings also enables new configurations for generating OAM beams that propagate at relative angles. For instance, by stacking a $q = 0$ grating and a $q \neq 0$ fork-dislocation grating in orthogonal orientations, followed by a sample, a single SANS measurement directly compares the scattering response of conventional and OAM-carrying neutrons within the same experimental geometry. This approach is promising for time-resolved SANS studies of dynamic or irreversible processes, where external parameters such as temperature or pressure are varied. Such experiments, examining phenomena like mechanical stress, particle growth, gel formation, or magnetic phase transitions, often involve structural evolution that cannot be reversed, sometimes necessitating multiple identical samples to ensure consistent conditions across repeated measurements~\cite{urban2021soft, C3CP50293G}. The techniques demonstrated here offer a practical solution: samples can be probed with multiple structured neutron states simultaneously, ensuring identical experimental conditions throughout the measurement. This may significantly benefit time-resolved SANS experiments targeting non-equilibrium or one-way transformations. Additionally, this technique enables direct experimental verification of whether OAM affects the neutron absorption cross section with polarized $^3$He~\cite{jach2022method}. Since the $^3$He cell depolarizes over time, implementing the $q=0$ and $q\neq0$ gratings in orthogonal orientations will enables a direct comparison between OAM-carrying and non-OAM neutrons. These techniques also open the door for generating more complex neutron waveforms, such as neutron Airy-vortex beams. This can be achieved by combining cubic and $q=3$ phase-gratings and post-selecting on the ($\pm~1$, $\pm~1$) diffraction orders. 

\section*{Acknowledgments}

This work was supported by the Canadian Excellence Research Chairs (CERC) program, the Natural Sciences and Engineering Research Council of Canada (NSERC) Discovery program, the NSERC Canada Graduate Scholarships programs (CGS-M and PGS-D), the Canada  First  Research  Excellence  Fund  (CFREF), and the US Department of Energy, Office of Nuclear Physics, under Interagency Agreement 89243019SSC000025. This work was also supported by the DOE Office of Science, Office of Basic Energy Sciences, in the program ``Quantum Horizons: QIS Research and Innovation for Nuclear Science'' through grant DE-SC0023695. A portion of this research used resources at the High Flux Isotope Reactor, a DOE Office of Science User Facility operated by the Oak Ridge National Laboratory.

\bibliographystyle{ieeetr}
\bibliography{mybib.bib}

\end{document}